\documentstyle[11pt,epsfig,aaspp4]{article}
\tighten 

\begin{document}

\newcommand{\cm}{cm$^{-2}$}
\newcommand{\qso}{Q0201+1120}

\newcommand{\lya}{Lyman~$\alpha$}
\newcommand{\lyb}{Lyman~$\beta$}
\newcommand{\za}{$z_{\rm abs}$}
\newcommand{\ze}{$z_{\rm em}$}
\newcommand{\cmtwo}{cm$^{-2}$}
\newcommand{\nhi}{$N$(H$^0$)}
\newcommand{\nzn}{$N$(Zn$^+$)}
\newcommand{\ncr}{$N$(Cr$^+$)}
\newcommand{\degpoint}{\mbox{$^\circ\mskip-7.0mu.\,$}}
\newcommand{\halpha}{\mbox{H$\alpha$}}
\newcommand{\hbeta}{\mbox{H$\beta$}}
\newcommand{\hgamma}{\mbox{H$\gamma$}}
\newcommand{\kms}{\,km~s$^{-1}$}      
\newcommand{\minpoint}{\mbox{$'\mskip-4.7mu.\mskip0.8mu$}}
\newcommand{\mv}{\mbox{$m_{_V}$}}
\newcommand{\Mv}{\mbox{$M_{_V}$}}
\newcommand{\peryr}{\mbox{$\>\rm yr^{-1}$}}
\newcommand{\secpoint}{\mbox{$''\mskip-7.6mu.\,$}}
\newcommand{\sqdeg}{\mbox{${\rm deg}^2$}}
\newcommand{\squig}{\sim\!\!}
\newcommand{\subsun}{\mbox{$_{\twelvesy\odot}$}}
\newcommand{\et}{et al.~}

\def\ltsima{$\; \buildrel < \over \sim \;$}
\def\simlt{\lower.5ex\hbox{\ltsima}}
\def\gtsima{$\; \buildrel > \over \sim \;$}
\def\simgt{\lower.5ex\hbox{\gtsima}}
\def\arcs{$''~$}
\def\arcm{$'~$}
\vspace*{0.1cm}

\title{
AN IMAGING AND SPECTROSCOPIC STUDY OF THE $z_{\rm abs} = 3.38639$
DAMPED Ly$\alpha$ SYSTEM IN Q0201+1120: CLUES TO STAR FORMATION
AT HIGH REDSHIFT \altaffilmark{1}}
\vspace{1cm}
\author{\sc Sara L. Ellison}
\affil{Institute of Astronomy, Madingley Road, Cambridge, CB3 0HA, UK and \\
European Southern Observatory, Casilla 19001, Santiago 19, Chile}
\author{\sc Max Pettini}
\affil{Institute of Astronomy, Madingley Road, Cambridge, CB3 0HA, UK}
\author{\sc Charles C. Steidel\altaffilmark{2} and Alice E. Shapley}
\affil{Palomar Observatory, Caltech 105--24, Pasadena, CA 91125}

\altaffiltext{1}{Based in large part on observations obtained 
at the W.M. Keck Observatory, which
is operated as a scientific partnership 
among the California Institute of Technology, the University of 
California, and NASA, and 
was made possible by the generous financial support of the 
W.M. Keck Foundation.}
\altaffiltext{2}{Packard Fellow}

\newpage
\begin{abstract}
We present the results of a series of imaging and spectroscopic
observations aimed at identifying and studying the galaxy responsible
for the $z_{\rm abs} = 3.38639$ damped \lya\ system in
the $z_{\rm em} = 3.61$ QSO  Q0201+1120.
We find that the DLA is part of a concentration of matter which 
includes at least four galaxies (probably many more)
over linear comoving dimensions greater than $5h^{-1}$~Mpc.
The absorber may be a $0.7 L^{\ast}$ galaxy at an impact
parameter of $15 h^{-1}$~kpc, but follow-up spectroscopy
is still required for positive identification.
The gas is turbulent, with many absorption
components distributed over $\sim 270$~km~s$^{-1}$
and a large spin temperature, $T_s \simgt 4000$~K.
The metallicity is relatively high for this redshift,
$Z_{\rm DLA} \simeq 1/20 Z_{\odot}$. From consideration
of the relative ratios of elements which have different
nucleosynthetic timescales, it would appear that
the last major episode of star formation in this DLA
occurred at $z \simgt 4.3$, more than $\sim 500$~Myr prior to the time
when we observe it. 
\end{abstract}


\keywords{cosmology:observations --- galaxies:abundances ---
galaxies:evolution --- quasars:absorption lines -- quasars:
individuals (Q0201+1120)} 

\section{INTRODUCTION}

The study of quasar absorption lines is a unique
tool with which to investigate the high redshift
universe on a variety of scales, ranging from the intergalactic
medium (IGM) to proto-galaxies. Since the detection of material
intercepting our line of sight to a given QSO is dependent only
on the column density of gas and the luminosity of the quasar,
this is a particularly powerful and sensitive technique for
probing the chemical composition and physical conditions in the
interstellar media (ISM) of high $z$ galaxies.  

Damped Lyman Alpha systems (DLAs) are the highest column density
($N$(H~I) $\geq 2 \times 10^{20}$ atoms \cm) absorbers identified
in QSO spectra and the major contributors to the census of
neutral gas available for star formation. 
DLAs are therefore widely believed to be the absorption
signatures of galaxies in early stages of evolution; the study of
their associated metal lines has provided us with the first
accurate measurements of the abundances of a wide variety of
chemical elements at high $z$ (Lu et al. 1996; Prochaska \& Wolfe
1999). Indeed, measuring the abundances in the ISM of DLAs not
only gives a direct view of the evolving chemistry of galaxies
over time, but is also a valuable testbed for supernova models
and their theoretical yields (e.g. Pettini et al. 2000).

However, the exact nature of DLAs is still a contentious issue.
Results from early DLA surveys (e.g. Wolfe et al. 1986; Lanzetta
et al. 1991) led to the proposal that these systems trace the
progenitors of present day spirals.  More recently, Prochaska \&
Wolfe (1997; 1998) have argued that this hypothesis is further
supported by the kinematics of the metal absorption lines which
they can best explain with large rotating disks with velocities
$v_{\rm rot} \simlt 200$~\kms\ (although see Ledoux et al. 1998).
On the other hand, the presence of such massive disks at $z \sim
3$ is in stark contrast with the predictions of hierarchical CDM
models (e.g. Kauffmann 1996) where galaxies at high redshifts are
relatively compact and underluminous.  In such models, the
kinematics of merging proto-galactic clumps can also reproduce
the shapes of the absorption lines (Haehnelt, Steinmetz, \& Rauch
1998; McDonald \& Miralda-Escud\'{e} 1999; Gardner et al. 2000). 
From the point of
view of their chemical enrichment, it was shown by Pettini et al.
(1997) that the distribution of DLA metallicities is displaced to
lower values relative to those those of the thin and thick disk
stellar populations of the Milky Way (see also Wolfe \&
Prochaska 1998).

At intermediate redshifts the picture is a little clearer. Direct
imaging with {\it HST} of galaxies associated with DLAs at $z
\simlt 1$ has revealed a range of morphologies with only a
minority of spirals (Le Brun et al. 1997; Rao \& Turnshek 1998;
Pettini et al. 2000). However, there are still concerns regarding
sample bias. An absorption system at $z \simlt 1.5$ can only be
recognised as a DLA with {\it HST} ultraviolet spectroscopy (or
$21\,$cm absorption, but such systems are rare). Thus QSOs behind
metal- and presumably dust-rich DLAs may well be
under-represented in existing surveys, given the bright magnitude
limit for spectroscopy imposed by the relatively small aperture
of {\it HST\/}. Since $B$-band luminosity and metallicity are
related (e.g. Kobulnicky \& Zaritsky 1999), {\it HST}-selected
DLAs are likely to be biased in favour of absorbers of low
luminosity and metallicity.
 
In order to make further progress on these issues, it is
necessary to bring together all the information provided by deep
imaging, kinematics, and chemical abundance studies into one
coherent picture. In this paper we explore this approach for a
high redshift DLA in the direction of the QSO Q0201+1120 which we
have targeted with a number of different observations. 

The compact, flat spectrum, radio source PKS~0201+113 with a
faint ($R = 19.5$) optical counterpart (Condon, Hicks, \& Jauncey
1977) was identified as a $z_{\rm em} = 3.61$ QSO by White,
Kinney, \& Becker (1993) who also reported the presence of a
strong DLA at $z_{\rm abs} = 3.3875$ with an 
estimated $N{\rm (H~I)} = 2.5 \times 10^{21}$~cm$^{-2}$.
This value of the neutral hydrogen column density is
sufficiently high to produce absorption in the redshifted 21~cm
line against the background radio source (de Bruyn, O'Dea, \&
Baum 1996; Briggs, Brinks, \& Wolfe 1997), 
although there is still some uncertainty about the
reality of these weak detections (Kanekar \& Chengalur 1997).
Here we present deep imaging, both from the ground and
with {\it HST}, of the field of Q0201+1120 aimed at detecting the
galaxy responsible for the DLA. In addition, we have obtained
intermediate and low resolution spectroscopy of the QSO and of a
number of faint sources in its proximity revealed by our images. High
resolution spectroscopy of the DLA is used to study its chemical
composition and velocity structure. All these data are
scrutinised for clues to the nature of the object producing the
damped \lya\ system. 

Unless otherwise stated, we use a 
$\Omega_M = 0.3$, $\Omega_{\Lambda} = 0.7$ cosmology;
$h$ is the Hubble constant in units
of 100~km~s$^{-1}$~Mpc$^{-1}$;
and all redshifts reported in this paper
are vacuum heliocentric. \\

\section{OBSERVATIONS}

Table 1 gives relevant details of the observations which we now
briefly discuss.

The top panel of Figure 1 shows a new spectrum of Q0201+1120
covering the wavelength range 3500--8000~\AA, obtained with the
Kast double spectrograph of the Shane 3.0~m telescope at Lick
Observatory. The spectral resolution is $\sim 4.5$~\AA, a factor
of $\sim 2$ better than the discovery spectrum by White et al.
(1993). We confirm both the redshift of the QSO (within the
errors of measurement) and the presence
of a strong damped \lya\ line near 5330~\AA; we defer precise
measurements of the absorption redshift and neutral hydrogen
column density to \S 2.3.1 and 3.1 respectively, where we present
the analysis of much higher resolution echelle observations.

\subsection{Imaging} 

The field of Q0201+1120 was imaged with the COSMIC camera (Kells
et al. 1998) at the prime focus of the Palomar 5~m Hale telescope
in the $U_n$, $G$, ${\cal R}$ photometric system designed by
Steidel \& Hamilton (1993) to detect primarily galaxies with
redshifts in the interval $2.7 \leq z \leq 3.4$\,. The thinned,
AR-coated Tektronix $2048 \times 2048$ CCD covers a $9.7 \times
9.7$~arcmin field sampled with a scale of 0.283 arcsec per pixel.
The seeing measured from the stacked images was $\sim
0.7-0.9$\,arcsec FWHM, and the $1 \sigma$ surface brightness
limits are $\sim 28.56,\,28.73,~{\rm and}~27.95\,$(AB) mag
arcsec$^{-2}$ in $U_n$, $G$, ${\cal R}$ respectively. However,
the non-negligible foreground extinction in this direction
(E($B-V$) = 0.147; Schlegel, Finkbeiner, \& Davis 1998) makes
this field less deep than most of the others observed in our
Lyman break survey. Figure 2 shows contour plots of the central
30 arcsec of the ${\cal R}$ image before (left) and after (right)
subtraction of the QSO image (achieved by appropriate scaling of
the point spread function (PSF) determined from nearby stars). 

We also acquired an image of this field
with the Wide Field Planetary Camera (WFPC2)
on {\it HST} through the F606W filter.
To produce the final image shown in Figure 3
we combined ten individual CCD frames by
`drizzling' onto a master output pixel grid;
the subtraction of the QSO used a careful
modelling of the PSF appropriate for the color of the QSO
and its position on the chip. Further details 
of the data reduction procedure can be found
in Pettini et al. (2000).

A number of faint sources are present near the QSO. The objects
of interest are labeled in Figures 2 and 3 and their relevant
parameters are collected in Table 2; the photometry includes
corrections for foreground reddening. Galaxies G1 and G3 are
unlikely to be associated with the damped \lya\ system since they
are both detected in $U_n$. The colours of G1 are consistent with
those of an early-type galaxy at $z \approx 0.5$, while G3 is
nearly flat-spectrum and is therefore likely to be at $z > 1$.
(Note that this galaxy is very faint (${\cal R} = 26.3$) so that
it can barely be seen in Figure 3 and is below the lowest contour
in Figure 2). Galaxies G2 and oM6, on the other hand, are both
$U_n$-drops and in principle could be at the redshifts of either
the QSO or the DLA. A possible caveat to these conclusions is
that the ground-based photometry of G1 and especially G2 is made
somewhat uncertain by blending with each other and with the QSO.
However, the WFPC2 F606W magnitudes of all the objects of
interest here are consistent with the ${\cal R}$ and ($G - {\cal
R}$) ground-based measurements in Table 2. Finally, note that the
source oM6, which appears elongated in Figure 2, is clearly
resolved by {\it HST} into two components, separated by
1.23~arcsec.

\subsection{Low resolution spectroscopy}

We obtained spectroscopic observations of Lyman
break galaxy (LBG) candidates in the field of Q0201+1120
with the Low Resolution Imaging
Spectrograph (LRIS---Oke et al. 1995) on the Keck~I telescope,
using different 
slit masks each covering approximately 
$4 \times 7$ arcmin. 
Exposure times were 2.5--3~hours per mask
in separate 1800~s integrations; with the 300~grooves~mm$^{-1}$
grating the resolution
of the spectra is $\sim 12.5$~\AA.
These observations resulted in 27 spectroscopically
confirmed LBGs with secure redshifts.
The object oM6 was among those
observed; its spectrum, reproduced in the lower
panel of Figure 1, exhibits a prominent 
\lya\ emission line at $z_{\rm em} = 3.645$.
In some of the exposures the two components
of oM6 are partially resolved, 
showing that the \lya\ emission line is stronger 
in the western component of the pair
in Figure 3.

The emission line appears to be spectrally resolved; its profile
is symmetric with FWHM~$\approx 450$~km~s$^{-1}$
after correction for the instrumental resolution.
We measure a line flux 
$f_{Ly\alpha} = {\rm (}8.7 \pm 0.3{\rm )}
\times 10^{-17}$~erg~s$^{-1}$~cm$^{-2}$
which corresponds to a \lya\ luminosity
$L_{Ly\alpha} \simeq 5.2 \times 10^{42} h^{-2}$~erg~s$^{-1}$ 
in our cosmology.
The difference in redshift between oM6 and Q0201+1120
is only $+465$~km~s$^{-1}$, well within the uncertainties 
(both systematic and random) in the determination of the 
QSO systemic redshift from the broad emission lines.
We conclude that the object oM6 is {\it not} the DLA absorber,
but it is associated with the QSO, from which it is separated by 
only $\sim 19 h^{-1}$~kpc. 

There are several other known examples of such QSO companions; the
properties of oM6 are similar, for instance, to those of the \lya\
emitter at $z = 4.695$ near the QSO BR~1202$-$0725 discussed by
Hu, McMahon, \& Egami (1996). What is particularly interesting,
in the case of oM6, is the clear detection of a stellar continuum
(see Figure 1). In at least this case, the companion appears to
be a galaxy, possibly in the process of merging with the QSO host
(Cen 2000), rather than simply a gas cloud lit up in \lya\ by
its proximity to the QSO. At $z \simeq 3$ our ${\cal R}$ filter
samples the rest-frame far-UV spectrum produced by the integrated
population of O and B stars and thus provides a measure of the
star-formation rate. With the transformation discussed by Pettini
et al. (1998), ${\cal R} = 23.97$ corresponds to SFR~$\simeq 12
h^{-2} \, M_{\odot}$~yr$^{-1}$ in our cosmology. Correcting for
dust reddening of the UV continuum as described in Adelberger \&
Steidel (2000), we find that ($G - {\cal R}{\rm )} = 1.34$
implies $E(B - V) = 0.079$ and an extinction at 1500~\AA\
$A_{1500} = 0.9\,$mag. Thus, the dust corrected SFR is $\simeq 28
h^{-2}\,M_{\odot}$~yr$^{-1}$. For comparison, assuming case B
recombination and Kennicutt's (1983) calibration, $L_{Ly\alpha}
\simeq 5.2 \times 10^{42} h^{-2}$~erg~s$^{-1}$ would imply only
SFR~$\simeq 5 h^{-2}\, M_{\odot}$~yr$^{-1}$---even if the \lya\
emission line were produced entirely by stellar photoionization,
which seems unlikely in this instance given the QSO proximity.
Evidently, there is significant suppression of \lya\ photons
within the stellar H~II regions of oM6, as is normally the case
in star-forming galaxies at high and low redshifts, or most of
the Lyman continuum photons escape the nebula.

Returning to Figure 1, it can also be seen that, although the
signal-to-noise ratio is not high, there appears to be a strong
absorption feature near 5300~\AA\ which we tentatively identify
as a damped \lya\ line  at $z_{\rm abs} = 3.364$ with $N$(H~I)$ =
4 \times 10^{20}$~cm$^{-2}$ (see inset). The difference in
redshift from the  $z_{\rm abs} = 3.3875$ DLA in the spectrum of
the QSO, corresponding to $\Delta v \simeq  -1600$~km~s$^{-1}$, 
is far too high for this to be the same absorbing galaxy covering
both sight-lines and we must therefore be dealing with two
separate  absorbers. Furthermore, two out of the other 26 LBGs
with spectroscopic redshifts in this field were found to be at
similar redshift as the two DLAs. 
They are c10 and oMD26 at $z= 3.366$ and 3.368 respectively; 
their spectra are shown in Figure 4 and relevant properties are
listed in Table 2. 
Thus we have found four objects at $z \simeq 3.37$ in this field, 
two detected in their stellar continua as
Lyman break galaxies and the other two identified via absorption
lines in the spectra of higher redshift sources.

Unfortunately our LRIS spectroscopy did not include object G2;
in any case it would be very difficult to separate it from the
QSO which is only 2.9~arcsec away and 6 magnitudes brighter.
If this is the DLA, then its impact parameter from the QSO
sight-line is $15 h^{-1}\,$kpc (proper
distance). Our estimated ${\cal R} \simeq 25.3$ would imply that
this galaxy has a rest-frame UV luminosity of about 
$0.7 L^{\ast}$, adopting ${\cal R} = 24.54$ = $L^{\ast}$
at $z \simeq 3$ from the LBG luminosity function
derived by Adelberger \& Steidel (2000) for our cosmology, 
and making the appropriate
luminosity-distance and $k$-corrections. If this galaxy turns out
not to be at the redshift of the DLA, then the absorber must be
fainter than ${\cal R} \approx 26.0$, or $\approx 0.4 L^{\ast}$
(with the usual caveats that we would miss brighter objects
if located {\it directly} underneath the QSO image, or if 
they are extended and of low surface brightness).

The remaining 24 LBGs 
are at redshifts
between 2.167 and 3.802 and include a galaxy, m32,
at $z= 3.645$ the
same redshift (within the errors) as Q0201+1120 and oM6.
Its spectrum is also shown in Figure 4; its
projected separation of 134~arcsec from 
the QSO sightline (Table 2)
corresponds to a comoving distance 
of $3.14 h^{-1}$~Mpc.

Summarizing the results of this section, our deep Palomar and
{\it HST} imaging data have shown that the strong DLA at $z_{\rm
abs} = 3.3875$ in front of Q0201+1120 is part of a concentration
of galaxies which includes at least four objects distributed over
linear comoving dimensions of more than $5 h^{-1}\,$Mpc. The absorber is
either a $\approx 0.7 L^{\ast}$ galaxy at an impact parameter of
$15 h^{-1}\,$kpc or is likely to be fainter than about $0.4
L^{\ast}$. We have also identified two galaxies which are
at the same redshift as the QSO; one of them is at a
projected separation of only $19 h^{-1}\,$kpc from the QSO.

\subsection{High Resolution Spectroscopy}

In order to study in more detail the physical properties of the 
gas giving rise to the DLA system in Q0201+1120, we used HIRES,
the echelle spectrograph on the Keck~I telescope
(Vogt 1992), to record the
spectrum of the QSO between 4500 and 6900~\AA. With the
0.86\arcs entrance slit the spectral resolution was
6~km~s$^{-1}$~FWHM. The observations extended over two nights;
the total integration time was 40\,500~s, typically made up of
4500~s long individual exposures. 

The data were reduced using Tom Barlow's HIRES reduction package
MAKEE\footnote{MAKEE is available from
http://www2.keck.hawaii.edu:3636/realpublic/inst/hires/makeewww}.
Having bias subtracted and flat-fielded the individual CCD
frames, MAKEE performs an optimally weighted extraction followed
by wavelength calibration (on a vacuum heliocentric scale) using
reference spectra of a Th-Ar hollow cathode lamp. Before
joining the echelle orders, the summed data array was mapped onto
a linear wavelength grid (0.03 \AA\ pixel$^{-1}$); the resultant
one-dimensional spectrum was then normalised to the 
QSO continuum determined by fitting splines to regions free of
absorption lines. The signal-to-noise ratio measured in the
continuum is S/N~$\simeq 10$--20; with $R = 19.5$, Q0201+1120 is
at the limit of the HIRES capability. The spectrum extends from
1026 to 1573~\AA\ in the rest frame of the $z_{\rm abs} = 3.3875$
DLA.

\subsubsection{The redshift of the DLA}

Figure 5 is a montage of selected metal absorption lines 
in the DLA. The high resolution of the echelle data 
reveal a complex absorption system with many 
individual components spread over a velocity interval of 
$\sim 270$~km~s$^{-1}$. The two components with the largest
optical depths are centered at redshifts 
$z_{\rm abs} = 3.38632$ and 3.38684; 
in the strongest transitions 
(O~I~$\lambda 1302$ and C~II~$\lambda 1334$)
they form a black absorption feature
centered at $z_{\rm abs} = 3.38639$ which we 
take as the reference redshift for the DLA.

None of these redshift values agree with those reported for 21~cm
absorption against the background continuum source. Three groups
have searched for the 21~cm line in this damped
system, with conflicting results. de Bruyn et al. (1996) using
the Westerbork synthesis radio telescope found an absorption line
at $z_{21} = 3.38699 \pm 0.00003$ with optical depth $\tau_{21} =
0.085 \pm 0.02$ and FWHM~$= 9 \pm 2$~km~s$^{-1}$. A detection was
also reported by Briggs, Brinks, \& Wolfe (1997) using the
Arecibo telescope, but the feature in their spectrum had
significantly different parameters: $z_{21} = 3.38716 \pm
0.00007$, $\tau_{21} = 0.037 \pm 0.008$ and FWHM~$= 23 \pm
5$~km~s$^{-1}$. Finally, Kanekar \& Chengalur (1997) could see no
absorption at all in data obtained with the Ooty radio telescope
and of comparable sensitivity to those of the other two studies.

Here we note that the two values of $z_{21}$ above correspond to
velocities of $+41$ and $+53$~km~s$^{-1}$ relative to $z_{\rm
abs} = 3.38639$ we have adopted as the zero point for the metal
absorption lines. Referring to Figure 5, it can be seen that
these velocities fall on the red edge of the main group of
absorption components, where the optical depth in the metal lines
is not high. It seems unlikely that there should be such a large
displacement between the optical depth of H~I and those of all
the other metal absorption lines (including neutral species such
as O~I and N~I) which, as can be seen from Figure 5, agree so
precisely in velocity among themselves. (Note that the damped
\lya\ line itself is too broad to constrain the absorption
redshift of H~I within useful limits). First, the redshift
differences between radio and optical data are far greater than
the systematic uncertainties affecting the wavelength scale of
either set of observations. Second, in two other high-$z$ DLAs,
toward the radio QSOs Q0454$-$020 and Q1331$+$170, the centroids
of the 21~cm and metal absorption lines differ by less than
10~km~s$^{-1}$ (Prochaska 1999; Prochaska \& Wolfe 1999). While
such velocity differences may reflect geometric differences in
the sight-lines probed by the radio and optical
observations---and may therefore offer the means to probe the
physical structure of the absorbers on small scales---the
40--50~km~s$^{-1}$ displacement found in Q0201$+$1120 is
surprisingly large, particularly given the compact nature of the
radio source. 
Thus, it seems to us that the redshift mismatch
we have uncovered adds to the existing doubts as to the reality
of the radio detections; clearly deeper 21~cm searches would be
highly desirable. For the moment, it would appear that a rather
high lower limit, $T_s \simgt 4000$~K, applies to the spin
temperature of the H~I gas (obtained by scaling the Kanekar \&
Chengalur (1997) limit to the revised value of $N$(H~I) we deduce
at \S 3.1 below).

We have examined the profiles of the metal absorption lines in
Figure 5 for clues as to the nature of the absorber.
Qualitatively, we do not see clear signs of the `edge leading
asymmetry' pattern which, according to Prochaska \& Wolfe (1997),
is best explained by a sight-line through a thick rotating disk.
Rather, the kinematics of the gas in this DLA appear to us to be
more chaotic, with components of varying strengths distributed
over a larger velocity interval than the likely projected
rotational speed of a $\approx L^{\ast}$ galaxy. The imaging data
do not help in this respect. Galaxy G2, if it is the absorber, is
very faint; our observations presumably only pick out the
central, high surface brightness, region and do not allow us to
make meaningful statements as to the overall morphology of this
object.\\

\section{ELEMENT ABUNDANCES}
In the next stage of the analysis we deduce ion column densities
by fitting the profiles of the absorption lines; this information
is then used to estimate the abundances of several elements in
the $z_{\rm abs} = 3.38639$~DLA.

\subsection{Neutral hydrogen column density}

Figure 6 shows the profile fit to the damped \lya\ line. The
\lya\ forest is very crowded at these redshifts leaving only
relatively few `continuum' windows; consequently the fit is not
as well constrained as is the case for DLAs in lower redshift
QSOs. Our best estimate is $N$(H~I)~$=$~($1.8 \pm 0.3$)$\times
10^{21}$~cm$^{-2}$\,. 
Not surprisingly, this value is lower than
$N{\rm (H~I)} = 2.5 \times 10^{21}$~cm$^{-2}$ deduced by White,
Kinney, \& Becker (1993) from consideration of the line width 
in their $\sim 10$~\AA\ resolution spectrum; at these high
redshifts blending with \lya\ forest lines tends to 
boost the apparent width of a damped line
with the result that the neutral hydrogen column density
is easily overestimated.

\subsection{Metal lines}

Although our HIRES spectrum covers the wavelength region
1026 to 1573~\AA\ in the rest frame of the DLA, many of 
the transitions of interest are blended with \lya\ forest lines.
We have fitted all the unblended lines (or unblended portions of
lines) with Voigt profiles using the 
VPFIT package\footnote{VPFIT is available at
http://www.ast.cam.ac.uk/\~rfc/vpfit.html}.
VPFIT determines the Doppler width ($b$), column density ($N$) and
redshift ($z$) of individual absorption components by minimizing 
the difference between observed and computed profiles; the number
of components is kept at the minimum required by the S/N of the
data.

Oscillator strengths for the transitions considered are from the
compilation by Morton (1991) with subsequent revisions as
summarised by Tripp, Lu, \& Savage (1996). For Ni~II~$\lambda
1370.1$ and $\lambda 1454.8$ we have used the recent
astrophysical and laboratory determinations by Zsarg\'{o} \&
Federman (1998) and Fedchak \& Lawler (1999).
However, the $f$-value for one of the Ni~II lines in our
spectrum, $\lambda 1317.2$, is not among those included in
these recent studies. By fitting simultaneously all three Ni~II
lines in our data with $f$(1317.2) as the free parameter, we
deduced $f$(1317.2) = 0.07, approximately a factor of two lower
than the theoretical value listed in Morton's (1991) compilation.
This correction is similar to those which Fedchack \& Lawler
(1999) measured for other Ni~II transitions, although data of
higher S/N than those considered here are clearly required to
refine this astrophysical determination.

From Figure 5 it can be seen that, as well as blending,
line saturation limits the number of ion species for which
reliable values of the column density can be determined. 
Specifically, O~I~$\lambda 1302.2$ and C~II~$\lambda
1334.5$ are too strong throughout
to be useful for abundance studies. 
In the case of Si~II, only $\lambda 1304.4$ is sufficiently
weak over part of the velocity range for VPFIT to provide a
reliable solution, but the components in question---between
100 and 240~km~s$^{-1}$---do not constitute the bulk of the 
absorbing gas.
This leaves only N~I, S~II, Fe~II, and Ni~II as relatively
unblended and unsaturated. We found that all the absorption lines
of these elements 
(N~I~$\lambda1200.0$ and $\lambda1134.7$ triplets; 
S~II~$\lambda 1253.8$; Fe~II~$\lambda 1122.0$ 
and $\lambda 1142.4$; and Ni~II~$\lambda 1454.8$, 
$\lambda 1370.1$, and $\lambda 1317.2$)
could be fitted satisfactorily with the same
set of VPFIT values ($b$ and $z$), leaving only the 
column density $N$ to vary between the different species.
The results are listed in column 2 of Table 3; typical errors
returned by the fitting procedure are $\pm 15$\%.

The fact that for some absorption lines the velocity range
100--240~km~s$^{-1}$ is not covered (see Figure 5) is not
a major concern because
the column density in this group of
components is less than 5\% of the total.
As for N~I, although the main components are saturated, we found
that the model parameters determined from the weaker unsaturated 
lines of S~II, Fe~II and Ni~II provided the best fit to N~I as
well.
The important point, for the discussion below, is that {\it it is not
possible to fit the N~I lines with significantly lower
values of column density than that listed in Table 3}.
Larger values are of course possible if the saturated N~I
lines hide extremely narrow components ($b < 2$~km~s$^{-1}$)
which are not present in the absorption lines from the first
ions--- an unlikely, but not impossible, scenario.

The species observed are the major ionization stages 
in H~I regions, so that their ratios relative to hydrogen
are a direct measure of the corresponding element abundances
(column 3 of Table 3).
The contributions from higher ions, which are generally small
in DLAs (Viegas 1995; Howk \& Sembach 1999), are likely to
be even less important in this case given that:
(a) the column
density of neutral gas is high,
and (b) the 
DLA is at a redshift where 
the intergalactic radiation field produced by QSOs is 
below its peak value at $z \sim 2.5$ (e.g. Madau 2000).
Comparison with the solar meteoritic abundance scale of 
Grevesse \& Sauval (1998) finally gives the relative abundances
listed in the last column of Table 3. 
Clearly, the gas producing the 
$z_{\rm abs} = 3.38639$ DLA is generally metal-poor,
with element abundances $Z \simlt 1/20 Z_{\odot}$.
This value is at the upper end of the range
measured in DLAs at $z > 3$ (Prochaska \& Wolfe 2000),
but somewhat lower than the values which 
seem to be typical of luminous
Lyman break galaxies at these redshifts 
(Pettini 2000 and references therein; Teplitz et al. 2000).

\section{DISCUSSION OF ELEMENT RATIOS}

In this section we analyse the relative abundances of the
elements observed for clues to the star-formation episodes which
enriched the DLA prior to $z \sim 3.4$\,. We follow the general
approach outlined by Pettini et al. (2000) and refer the
interested reader to that work for a more detailed discussion of
the underlying ideas. The basic premise is that the elements in
question have different nucleosynthetic histories so that their
relative abundances are not expected to remain constant following
a burst of star formation. Rather, the element ratios will change
as stars of different masses (and therefore lifetimes) evolve off
the main sequence and contribute to the chemical enrichment of
the gas. As we shall see, ours is a rather ambitious goal given
the uncertainties in the nucleosynthetic processes involved, in
the stellar lifetimes, and in our own abundance measurements for this
DLA; consequently, we can at most hope for clues, rather than
definite conclusions.

\subsection{Overabundance of the $\alpha$-elements} 

One of the cornerstones of galactic chemical evolution models is
the overabundance of the $\alpha$-elements relative to Fe seen in
Galactic metal-poor stars (where Fe is taken as the measure
of metallicity).
This effect is generally interpreted as reflecting the
delayed production of $\approx 2/3$ of the iron by Type~Ia
supernovae, but in recent years this simple scenario has begun to
show its limitations. First, there appear to be real differences
between different elements within the $\alpha$-capture group
(e.g. Chen et al. 2000; Prochaska et al. 2000). The most obvious
of these is the finding that the ratio [O/Fe] shows a steady
increases with decreasing [Fe/H] and reaches [O/Fe]~$\sim +1.0$
at [Fe/H]~$\sim -3$. In contrast, most (but not all? --- see
below) of the other $\alpha$-elements seem to have a constant
[$\alpha$/Fe]~$\sim +0.5$ between [Fe/H]~$\sim -1$ and $-3$
(Takeda et al. 2000 and references therein). Second, as the
progenitors of Type Ia supernovae have yet to be positively
identified, the timescale for the release of Fe is uncertain
(see, for example,  the comprehensive review by Nomoto et al.
1999). It is customary to assume a $\sim 1$~Gyr delay between the
production of O and other $\alpha$-elements in Type~II SN on the
one hand, and
the release of the bulk of the Fe-peak elements by Type Ia SN
on the other, 
but it is possible that the latter follow the former on much
shorter intervals. For example, Reg\H{o}s et al. (2000) have
recently pointed out that the peak in the rate of Type Ia SN
formed by the edge-lit detonation of CO white dwarfs is reached
less than 200~Myr after the onset of star formation. This would
drastically reduce the evolutionary timescales implied by the
$\alpha$-elements overabundances in halo and thick-disk stars of
the Milky Way.

Returning to the $z_{\rm abs} = 3.38639$ DLA, the only
$\alpha$-element we cover in our spectrum is sulphur.
Its behavior at low [Fe/H] is not as well documented
as that of other $\alpha$-elements because the sulphur
abundance is determined from only one, high excitation, line
in the far red (S~I~$\lambda 8694$). The studies by 
Francois (1988) and most recently Takeda et al. (2000)
found S to track O, while Prochaska et al. (2000)
found a large scatter about [S/Fe]~$\approx 0$ 
(i.e. no enhancement of sulphur)
in thick disk stars with [Fe/H]~$= -0.4$ to $-0.6$\,.
All these authors stress the importance of carrying
out more extensive surveys of the abundance of sulphur
in the future.

From the last column of Table 3 it can be seen 
that we find [S/Fe]~$\simeq +0.16$ at [Fe/H]~$ = -1.41$.
This represents at most a mild S overabundance, given 
our error of $\pm 0.1$ in the log of the ratio of any two
elements and the possibility that Fe, unlike S, may be partly
depleted onto dust. For comparison, Francois (1988)
finds [S/Fe]~$\simeq +0.6$ in Milky Way stars with
[Fe/H]~$= -1.4$\,. In the context of the discussion 
above, we would conclude that there has been sufficient
time, since the last episode of star formation
which enriched this damped \lya\ system in metals,
for Fe to build up to a near-solar value relative to S.
The lack of a pronounced enhancement of the 
$\alpha$-elements now seems common to many 
low-metallicity DLAs (e.g. Pettini et al. 2000; 
Centuri\'{o}n et al. 2000) and has been interpreted
as evidence for generally low rates of star formation.

\subsection{N/$\alpha$}
Nitrogen is another element at our disposal which may be
used to date the chemical enrichment process.
At low metallicities the main source of N is thought to be 
primary nucleosynthesis in intermediate mass stars
($2 - 5 M_{\odot}$)
with a time delay of 250--500~Myr relative to the 
near-instantaneous release of O after a burst of
star formation (Henry, Edmunds, \& K\"{o}ppen 2000;
Lattanzio et al. 2000). 
The [N/$\alpha$] ratio in DLAs exhibits a 
much larger scatter than any other element ratio not 
involving N (Lu, Sargent, \& Barlow  1998; Centuri\'{o}n et al. 1998),
with values spanning the range between secondary and primary
nitrogen production.
Pettini, Lipman, \& Hunstead (1995) proposed that this scatter
is most naturally explained as the result of the delayed release
of primary N, although there are dissenting viewpoints
(Izotov \& Thuan 1999).

From the last column of Table 3 it can be seen 
that in the $z_{\rm abs} = 3.38639$ DLA we find
N to be less abundant than S by a factor of 4,
that is [N/S]~$ = -0.6$\,. With the assumption that 
[S/O] = 0, this corresponds to
(N/O)~$ = -1.51$ at (O/H)~$+ 12 = 7.58$\footnote{We apologise
to the reader for this sudden change of notation, but nearly
all published measurements of the N and O abundances are
in these units. (N/O) and (O/H) are the logarithmic values
of the element ratios by number, without reference to the 
solar values.}. Referring to Figure 5 of Pettini et al. (1995),
it can be seen that (N/O)~$ = -1.51$ is a `high' value, 
in the sense that it is 
close to that expected following the release of primary nitrogen.
It is at the upper end of the range measured in DLAs 
and in good agreement with the typical (N/O) in
blue compact and H~II
galaxies with (O/H)~$+ 12 \simeq 7.6$ (e.g. see Figure 1b of
Henry et al. 2000). 

In concluding this section, it would seem that, 
within the current understanding of their time evolution,
both the [$\alpha$/Fe] and the [N/$\alpha$] ratios
measured in this DLA point to the last major episode
of metal enrichment having occurred several hundred Myr
prior to the redshift at which we observe the DLA.
Adopting 500~Myr as the time lag for both N and Fe
production, places the last major episode of star
formation in this galaxy at $z > 4.3$ 
($H_0 = 65$~km~s$^{-1}$~Mpc$^{-1}$). This
conclusion is not at odds with what we 
already know about star formation at high redshifts,
given that there seems to be no change in the
luminosity function of star forming
galaxies between $z = 3$ and 4.5 (Steidel et al. 1999),
and that many LBGs have stellar populations older than
500~Myr (Shapley et al., in preparation).

As emphasized above, this type of analysis is based on many
assumptions whose validity remains to be verified;
ultimately it will only be possible to establish 
whether element ratios
give a self-consistent picture by conducting
a full study of a large body of high quality
measurements. However, we note that at least one other 
well-studied high-$z$ DLA has similar chemical properties
to those deduced here. From VLT/UVES observations
of the $z_{\rm abs} = 3.3901$ DLA in Q0000$-$2620
Molaro et al. (2000) deduce [$\alpha$/Fe-peak]~$\simeq +0.2$
and (N/O)~$ = -1.69$ at (O/H)~$+ 12 = 7.01$\,.
So this may well be another example of a DLA galaxy with low, 
or episodic, star formation which started before $z = 4.3$\,.\\

\section{Summary and Conclusions}

In this paper we have presented imaging and spectroscopic observations
aimed primarily at investigating the nature of the absorber
giving rise to the strong damped \lya\ system at $z_{\rm abs} = 3.38639$
in front of Q0201+1120. Our main findings can be summarised as follows.

1. The DLA is part of a concentration of matter which includes at
least four galaxies at $z \simeq 3.37$ over transverse dimensions
of at least $5 h^{-1}$~Mpc (comoving). Two of the galaxies are
seen directly via their UV stellar continua, while two others
(including the DLA) are detected in absorption against higher
redshift sources. Since we have so far obtained spectra for only
about one quarter of the $U$-drop candidates in this field, it is
likely that the structure includes other objects at $z \simeq
3.37$.

2. We have found a promising candidate for the absorber in a
photometric Lyman break object only 2.9~arcsec away from the QSO
sight-line. If confirmed by future spectroscopy to be at the 
absorption redshift, then the galaxy associated with this DLA has
a UV luminosity of about $0.7 L^{\ast}$ and a linear extent of at
least $15 h^{-1}\,$kpc. Otherwise the absorber is likely to be
fainter than about $0.4 L^{\ast}$.

3. The H~I gas producing the damped system is highly turbulent,
with a spin temperature $T_s \simgt 4000$~K and complex 
absorption line profiles, consisting of many discrete components
spanning $\sim 270$~km~s$^{-1}$. Neither is what one may have expected
from a cold, rotationally supported, disk.

4. The DLA has a metallicity $Z \simeq 1/20 Z_{\odot}$, at the upper end
of the distribution of values of $Z_{\rm DLA}$ at these redshifts,
but lower than those deduced for other luminous Lyman break
galaxies, although the data here are still very sparse.
Further, it exhibits no
marked overabundance of the $\alpha$-elements (in this case S) relative to
Fe, and a relatively high (N/$\alpha$) ratio for its low metallicity.
Both element ratios can be understood if there has been a relatively
quiescent interval, lasting more than $\sim 500$~Myr, since the last
major burst of star formation in this galaxy, 
allowing the delayed release of Fe from Type~Ia SN 
and of N from AGB stars of intermediate mass.
If this reasoning is correct, 
the star formation episode responsible
for producing the heavy elements we see occurred
at $z \simgt 4.3$\,. The chemical properties of this 
absorber are similar to those of another high-$z$ DLA,
at $z_{\rm abs}  = 3.3901$ in Q0000$-$2620, 
which has been extensively studied 
with both the Keck and VLT echelle spectrographs.

Firm conclusions on the nature of DLAs at high redshifts are 
hampered by the lack of follow-up spectroscopy of galaxy G2.
If this is indeed the absorber, it would be only the
second case where a high-$z$ DLA has been identified with a Lyman
break galaxy, after the $z_{\rm abs} = 3.151$ system in
Q2233+1310 (Steidel, Hamilton, \& Pettini 1995; Djorgowski et al.
1996), although others have been detected in \lya\ emission (e.g.
Fynbo, Burud, \& M{\o}ller 2000 and references therein). In
general our deep Lyman break imaging has shown that the galaxies
producing DLAs at $z \simgt 3$ must be significantly
sub-$L^{\ast}$ in their stellar continua (e.g. Steidel et al.
1995; Steidel et al. 1998) and that detections such as this one
are the exception rather than the rule. 

We conclude that, although the statistics are still very limited,
DLAs seem to sample a wide range of the luminosity function of
galaxies at $z \simeq 3$. QSO absorption line spectroscopy still
gives us the most precise measurements of many important physical
properties of high redshift gas. However, 
only by combining it with deep imaging
will it be possible to realise its full potential for unravelling the
nature and evolution of galaxies at early times.\\

We are grateful to the staff of the Palomar and Keck
Observatories for their competent technical assistance. Our
collaborators in the LBG survey project, Kurt Adelberger, Mark
Dickinson, Mauro Giavalisco, and Mindy Kellogg, generously helped
with various aspects of the data acquisition and reduction.
C.C.S. acknowledges support from the National Science Foundation
through grant AST 95-96229 and from the David and Lucile Packard
Foundation.

\newpage

%
%

\begin{deluxetable}{lllcr}
\tablecaption{Summary of Observations}
\tablehead{
\colhead{Object/Field} & \colhead{Telescope} & \colhead{Instrument} 
& \colhead{Filter/Grating} & \colhead{Int Time (s)}} 
\startdata
Q0201$+$1120 & Hale 5m     & COSMIC & $U_n$               & 27\,000  \nl
Q0201$+$1120 & Hale 5m     & COSMIC & $G$                 &  7\,200  \nl
Q0201$+$1120 & Hale 5m     & COSMIC & ${\cal R}$          &  3\,600  \nl
\nl
Q0201$+$1120 & {\it HST}   & WFPC2  & F606W               & 10\,800  \nl   
\nl
Q0201$+$1120 & Lick 3m     & Kast   & 600/4340 \& 600/7500  & 14\,400  \nl
Q0201$+$1120 & Keck~I 10m  & HIRES  &                     & 40\,500  \nl
\nl
Galaxy oM6   & Keck~I 10m  & LRIS   & 300/5000            & 10\,800  \nl
Galaxy m32   & Keck~I 10m  & LRIS   & 300/5000            & 10\,800  \nl
Galaxy c10   & Keck~I 10m  & LRIS   & 300/5000            & 10\,800  \nl
Galaxy oMD26 & Keck~I 10m  & LRIS   & 300/5000            & 10\,800  \nl
\enddata
\end{deluxetable}

\begin{deluxetable}{lrrrrrrr}
\tablecaption{Parameters of Sources in the Field of Q0201+1120}
\tablehead{
\colhead{Name} & \colhead {$\Delta$RA (\arcs)\tablenotemark{a}} 
& \colhead{$\Delta$Decl. (\arcs)\tablenotemark{a}}
& \colhead{${\cal R}$} & \colhead{$G - {\cal R}$} & \colhead{$U_n - G$}
& \colhead {Redshift} & \colhead{$\Delta r$ (Mpc)\tablenotemark{b}}
}
\startdata
Q0201$+$1120 & 0.0      & 0.0     & 19.23 & 0.93 & $\geq5.20$\tablenotemark{c} & 3.638 & 0.0 \nl
G1           & +0.9     & +0.4    & 22.23 & 1.48 & 1.63 & \ldots & \ldots   \nl
G2           & +2.8     & $-0.9$  & 25.3:\tablenotemark{d} & 0.5:\tablenotemark{d} & $> 1.7:$\tablenotemark{d}& \ldots & \ldots   \nl 
G3           & +3.4     & $-2.7$  & 26.3  & 0.2  & 0.3  & \ldots & \ldots \nl  
oM6          & +1.3     & $-3.5$  & 23.97 & 1.34 & $> 2.62$   & 3.645 & 0.085\nl
m32          & +113.5   & +71.5   & 25.21 & 0.79 & $> 1.99$   & 3.645 & \ldots   \nl
c10          & $-116.6$ & +166.1  & 24.07 & 0.70 & $> 2.68$                & 3.366 & 4.61 \nl
oMD26        & $-98.0$  & $+204.7$ & 24.72 & 0.83 &$\geq1.72$\tablenotemark{c}                    & 3.368 & 5.15 \nl
\enddata
\tablenotetext{a}{Relative to Q0201+1120.}
\tablenotetext{b}{Comoving distance from Q0201+1120 sightline at $z = 3.37$ 
($\Omega_{M} = 0.3$, $\Omega_{\Lambda} = 0.7$, $h = 1$).}
\tablenotetext{c}{Formally detected in $U_n$ at the $1 \sigma$ level.}
\tablenotetext{d}{The photometry of G2 is uncertain because of
its proximity to G1 and the QSO.}
\end{deluxetable}{}

\begin{deluxetable}{lcccc}
\tablecaption{Ion Column Densities and Element Abundances in the 
$z_{\rm abs} = 3.38639$ DLA}
\tablehead{
\colhead{$X^i$} & \colhead{log $N$($X^i$)\tablenotemark{a}} & 
\colhead{log $N$($X^i$)/$N$(H$^0$)} 
& \colhead{log ($X$/H)$_{\odot}$\tablenotemark{b}} 
& \colhead{[$X$/H]\tablenotemark{c}}
} 
\startdata
H$^0$  &   21.26   & \ldots   & \ldots  & \ldots  \nl
N$^0$  &   15.33   & $-5.93$  & $-4.08$ & $-1.85$ \nl
Si$^+$ &   $>14.00$\tablenotemark{d} & \ldots & \ldots & \ldots \nl
S$^+$  &   15.21   & $-6.05$  & $-4.80$ & $-1.25$ \nl
Fe$^+$ &   15.35   & $-5.91$  & $-4.50$ & $-1.41$ \nl
Ni$^+$ &   13.84   & $-7.42$  & $-5.75$ & $-1.67$ \nl
\enddata
\tablenotetext{a}{Log of the column density of ion $X^i$; typical
errors are $\pm 15$\%.}
\tablenotetext{b}{Solar meteoritic abundance scale from Grevesse
\& Sauval (1998).}
\tablenotetext{c}{[$X$/H] = log ($X$/H) $-$ log ($X$/H)$_{\odot}$.}
\tablenotetext{d}{This value refers only to
gas in the velocity interval 100---240~km~s$^{-1}$ (relative to
$z_{\rm abs} = 3.38639$) which accounts for only a small fraction
of the total column density.}
\end{deluxetable}{}

%
%

\begin{figure}
\hspace*{-0.7cm}
\centerline{\rotatebox{270}{\resizebox{14.5cm}{!}
{\includegraphics{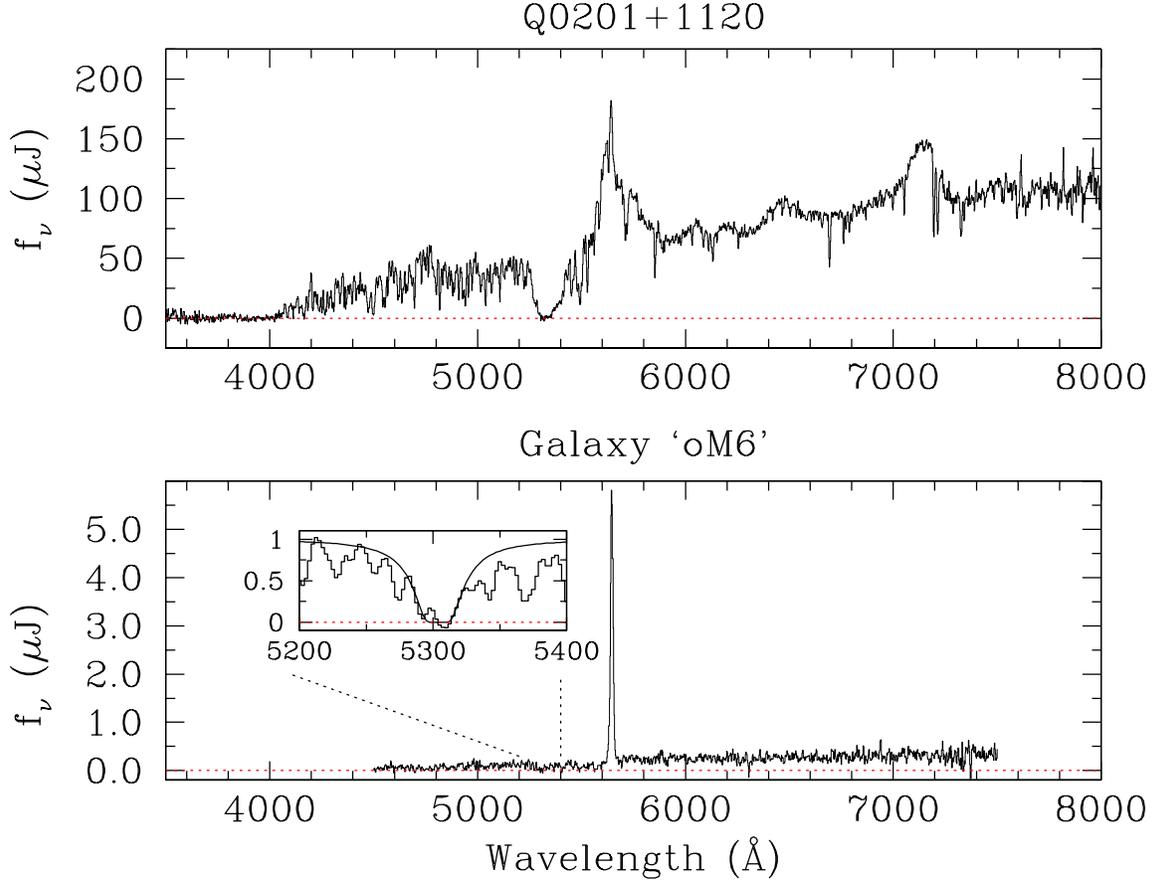}}}}
\caption{{\it Upper panel:} Lick spectrum of Q0201+1120; the resolution is 
$\sim 4.5$~\AA. {\it Lower panel:} LRIS spectrum of galaxy oM6; the 
resolution is $\sim 12$~\AA. See Table 1 for additional
details of the observations. The inset shows the profile fit
to the absorption feature near 5300~\AA\ 
which we tentatively identify as damped \lya\ at $z_{\rm abs} = 3.364$
with $N$(H~I)$ = 4 \times 10^{20}$~cm$^{-2}$. 
}
\end{figure}

\begin{figure}
\centerline{\rotatebox{90}{\resizebox{10cm}{!}{\includegraphics{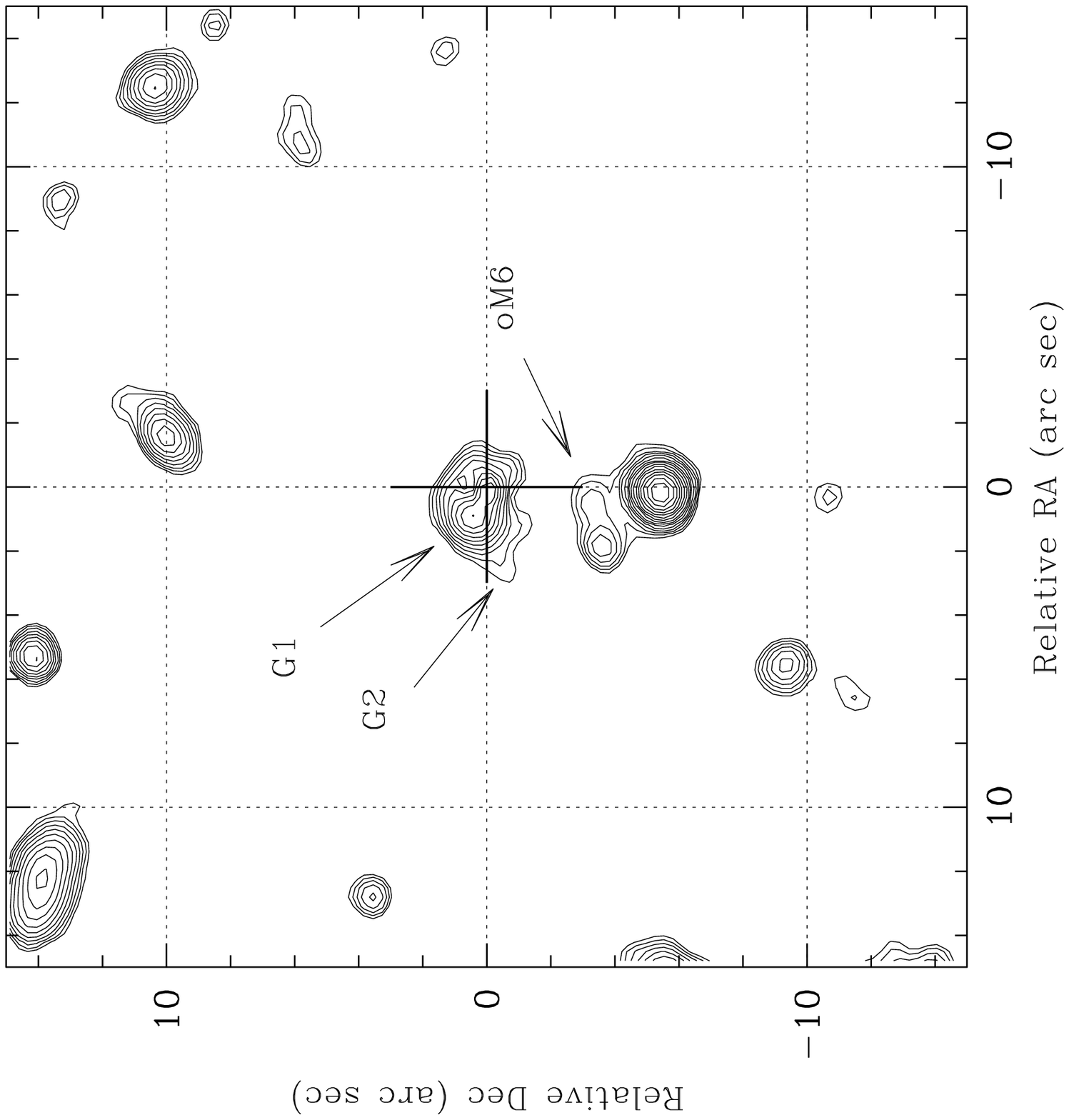}}}}
\centerline{\rotatebox{90}{\resizebox{10cm}{!}{\includegraphics{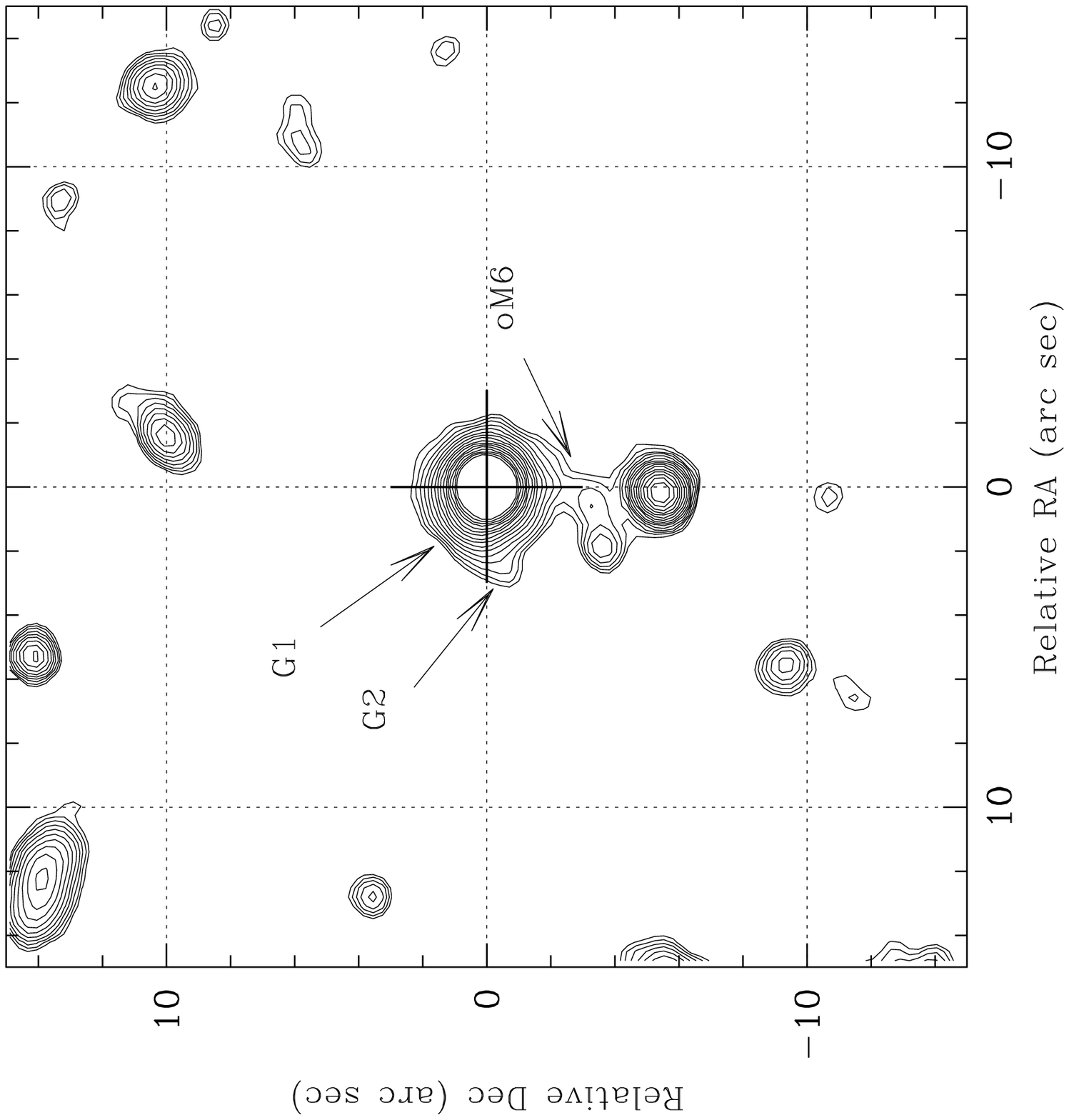}}}}
\caption{Contour plots of the central 30 arcsec of the Palomar
${\cal R}$-band image of the field of Q0201+1120 before (left)
and after (right) subtraction of the QSO image.}
\end{figure}

\begin{figure}
\centerline{\resizebox{8cm}{!}{\includegraphics{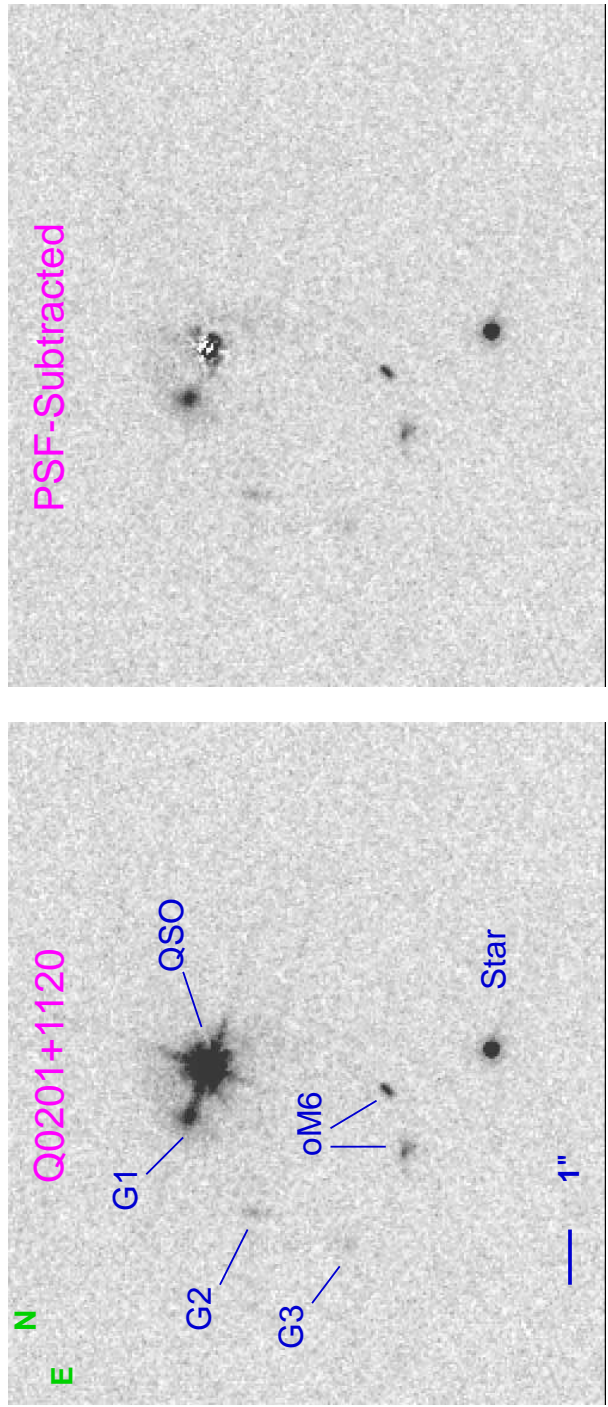}}}
\vspace{1cm}
\caption{F606W WFPC2 image of the field of Q0201+1120 before (left)
and after (right) subtraction of the QSO image.}
\end{figure}

\begin{figure}
\vspace{-2cm}
\centerline{\resizebox{18cm}{!}{\includegraphics{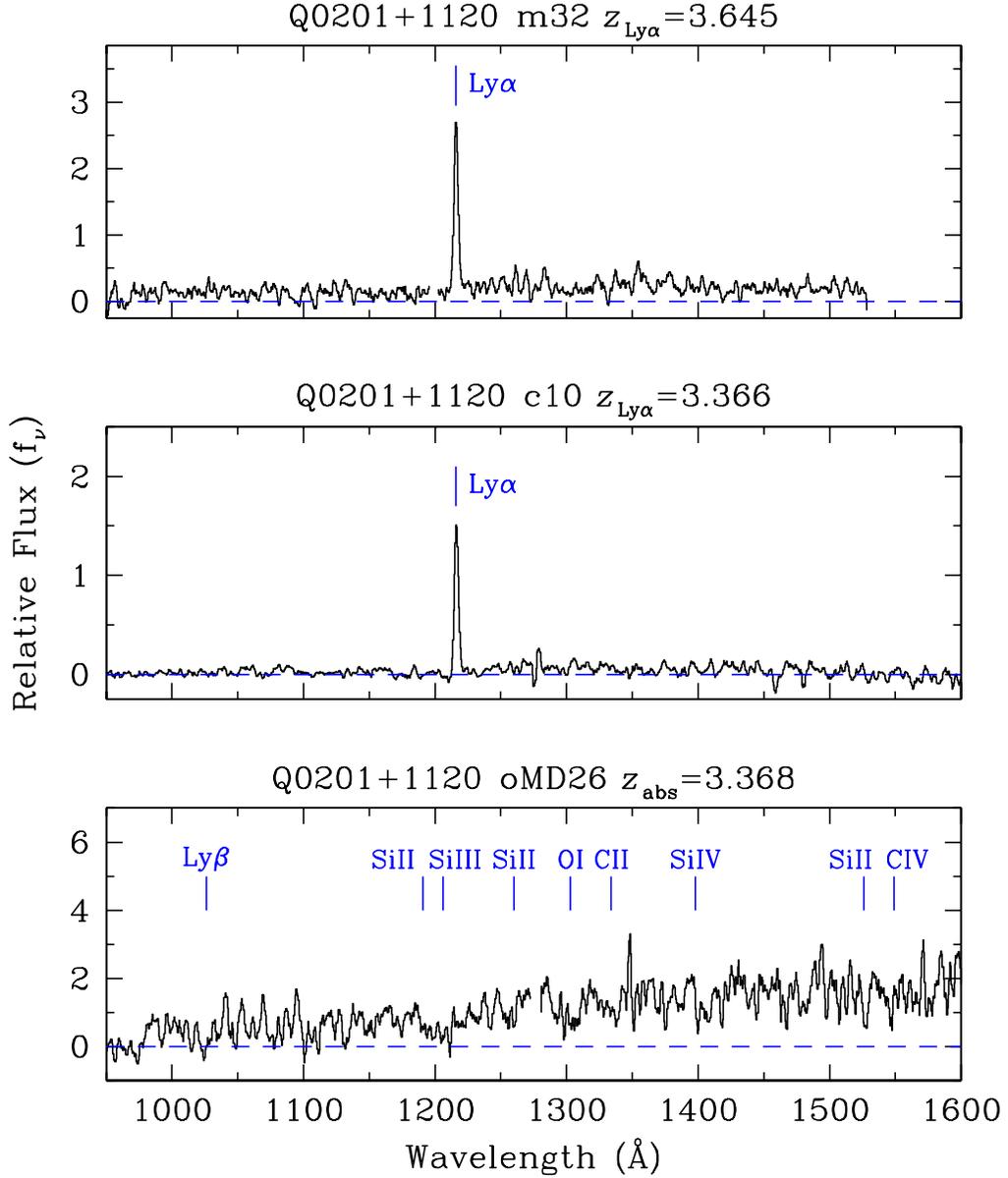}}}
\vspace{-3cm}
\caption{LRIS spectra of Lyman break galaxies m32, c10 and oMD26
reduced to the respective rest frames.
The redshifts of m32 and c10 are determined from their 
prominent \lya\ 
emission lines; that of oMD26 is from the strongest interstellar
absorption lines, as indicated. The redshift of m32 is close
to those of Q0201+1120 and oM6, while c10 and oMD26 are at
similar redshifts as the damped \lya\ system in Q0201+1120
(see Table 2).
}
\end{figure}

\begin{figure}
\centerline{\resizebox{15cm}{!}{\includegraphics{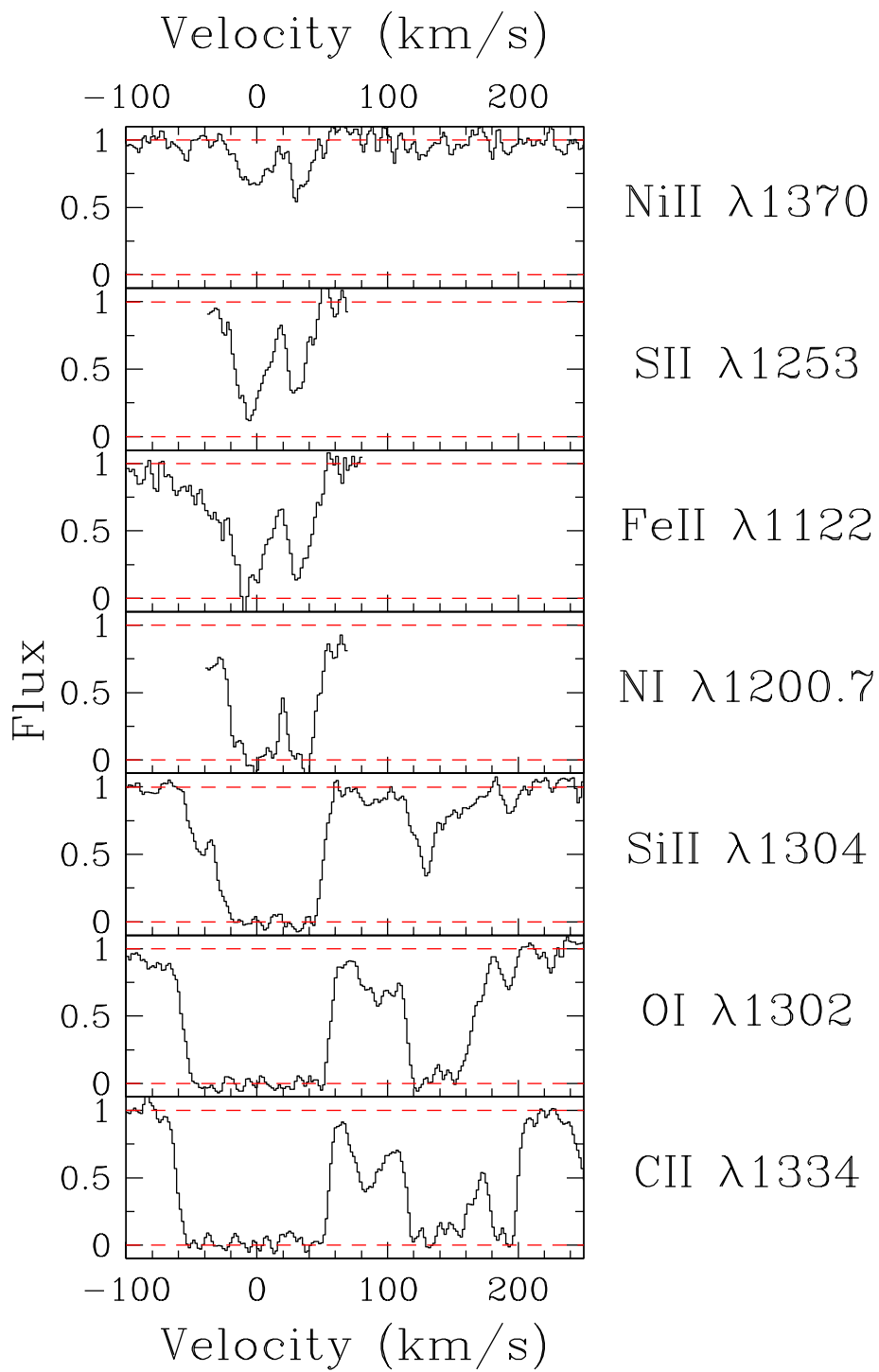}}}
\caption{\label{abund}Normalized profiles of selected 
metal absorption lines in the damped \lya\ system
at $z_{\rm abs} = 3.38639$; the velocity scale is relative 
to this redshift.
}
\end{figure}

\begin{figure}
\hspace*{-0.7cm}
\centerline{\rotatebox{270}{\resizebox{15cm}{!}
{\includegraphics{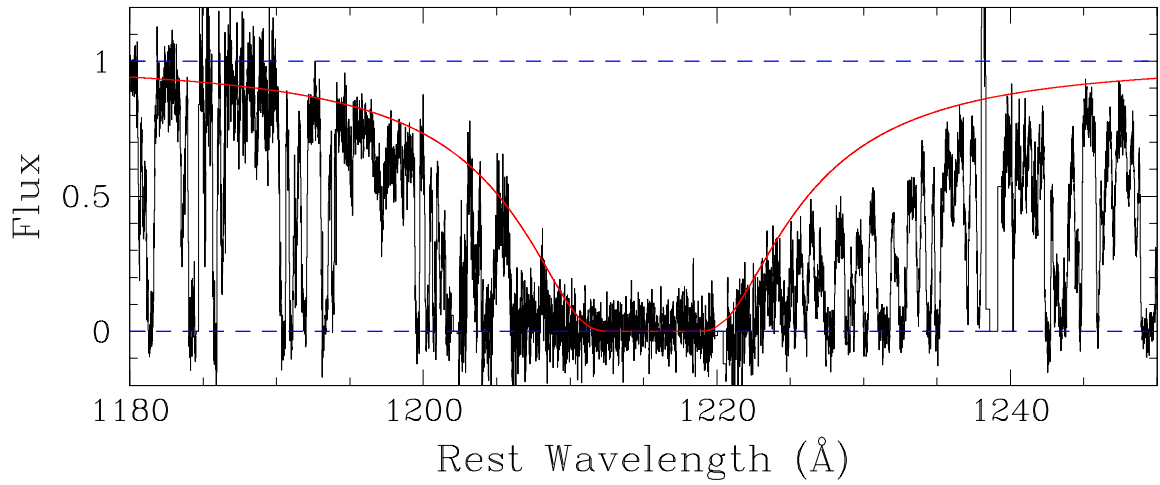}}}}
\caption{\label{hi_fit}
Portion of the HIRES spectrum of Q0210+1120 showing the 
damped \lya\ line at $z_{\rm abs} = 3.38639$.
The profile fit shown is for a column density 
$N$(H~I)\,$ = 1.8 \times 10^{21}$~cm$^{-2}$.}
\end{figure}

\end{document}